\documentstyle[12pt]{article}
\title{Decay of metastable multicomponent mixture}
\author{Victor Kurasov
\thanks{
Free University of Berlin, Physics faculty, TKM,
e-mail: kourassov@vaxneu.dnet.fu-berlin.de;
Saint-Petersburg University, Physics faculty,
Department of computational physics,
e-mail: kurasov@onti.phys.lgu.spb.su}
}
\date{       }

\begin{document}
\maketitle

\begin{abstract}
The kinetics of the
multicomponent condensation after the instantaneous creation
of the metastable state is described analytically. Manydimensional
problem is reduced to the one-dimensional case. All the main characteristics
of the process are obtained analytically with the help of iteration procedure.
As the result the analytical theory for the decay of the metastable phase
in completely constructed for vapor with many components.
\end{abstract}

\pagebreak

\section{Introduction}

The present paper is devoted to
the constuction of the kinetics of the condensation of multicomponent mixture
 after the
instantaneous creation of supersaturation in the system.

The theory presented here is based on the classical theory of nucleation.
The basic ideas of the classical theory of nucleation are created by Volmer
[1],
Becker and Doering [2], Zeldovitch [3], Frenkel [4], Kramers [5]. Some
reconsiderations
were made by Lothe and Pound [6]. The modern state of the
nucleation theory can be characterized by the contributions of Reiss [7],
Reiss, Katz and Cohen [8], Feder, Russel, Lothe and Pound [9],
Hung, Katz and Krasnopoler
[10], Reiss, Tabazadeh and Talbot [11] and others.

The creation of  stationary classical nucleation theory
allows to study kinetical problems of nucleation. Among
the first publications devoted to this topic one must stress
 the contributions by Wakeshima [12] and Reiser [13].
The theory of the homogeneous decay of the metastable
phase was created by Grinin, Kuni and Kabanov [14].
The theory of homogeneous condensation dynamical conditions was examined in
 [15],[16]. The theory of heterogeneous condensation was investigated in [17],
[18].
Binary condensation was
investigated by Reiss in [19],and  later by  Stauffer in [20]. The rate of
binary
nucleation
was obtained in [19] and was corrected in [20]. The stationary theory of binary
condensation with the help of Lorentz transformation was investigated in [21].

All above cited publications form the base for the theory of the decay of
metastable mixture presented here. This publication can be regarded as
some modification of ideas presented in [22]-[24].

Some concrete model has to be accepted in order to give us an opportunity to
fulfill detailed calculations. This model includes the following positions:

System is homogeneous in space.

The regime  of substance exchange between an embryo and environment is free
molecular
one.

Thermal effects of condensation are neglected.

All these assumptions can be overcome by modification of the theory and aren't
observed here due to the lack of the volume of the publication.

Coalescence (Ostwald ripenning)
 isn't analysed here. An asymptotical expansions are constructed
in [25],[26].

We shall choose the system of unit volume and measure all energy-like values in
thermal units.

\section{Embryos free energy}

For free energy of near-critical embryo of
liquid phase which consists of $\nu_{i}$
molecules of $i$ component for all $n$ components of solution
we have
\begin{equation}
F(\{\nu_{i}\}_{i=1,n}) = a(\sum_{i} v_{i} \nu_{i})^{2/3}
 -\sum_{i} b_{i}\nu_{i}
\end{equation}
where
\begin{equation}
a = 4 \pi \sigma (\frac{3}{4 \pi})^{2/3}
\end{equation}
$v$ is the fraction of partial volumes of the molecules of $a$ and $b$
components
in liquid phase, $\sigma$ is surface tension expressed in units $k_{b}T$,
 $k_{b}$ is Boltzman constant, $T$ is absolute  temperature.
Let us illustrate the method by using the following
 expressions for $b_{a}$ and $b_{b}$ from the theory of liquids.
\begin{equation}
b_{i} = \ln \frac{n_{i}}{n_{ii}} - \ln[\mu_{i}
f_{i}(\mu_{i})]
\ \ \ \ \ \mu_{i}= \frac{\nu_{i}}{\sum_{j} \nu_{j}}
\end{equation}
Here $n_{a}$ (or $n_{b}$) is the molecules number
density of component $a$ (or $b$) in the vapor;
$n_{aa}$ (or $n_{bb}$) is the molecules  number
density of the vapor saturated over plane surface.
The values of $f_{i}$ are called the coefficients of activity.
When $f_{i}\equiv 1$ we have an ideal solution.
The conditions of applicability of equation (1) are the following ones
\begin{equation} \label{4}
\nu_{i} \gg 1
\end{equation}
Inequality (\ref{4}) extracts so called cappilarity approximation in binary
nucleation.

\section{Droplets  growth}

Let us  use the word "droplets" to define the super-critical embryos
moving on the plane of sizes irreversibly.
In order to obtain the
velocity of their growth we take into account that due to free-molecular
regime of exchange the flow $W^{+}_{a}$ of $a$ component on the molecule is
equal to
\begin{equation}
W^{+}_{i}(\{\nu_{i}\}) = \frac{(\zeta_{i}+1)}{\tau_{i}}
(\sum_{j} v_{l\ j}\nu_{j})^{2/3}
\end{equation}
where $\zeta_{i}$ is the supersaturation of the $i$ component
 defined as
\begin{equation}
\zeta_{i} = \frac{n_{i}-n_{ii}}{n_{ii}}
\end{equation}
and $\tau_{i}$ is the characteristic time between collisions of
fixed molecule of component $a$ with other molecules of component $a$ in the
vapor
\begin{equation}
\tau_{i} \sim [  n_{ii} v_{t\ i}]^{-1}
\end{equation}
$v_{t\ i}$ is the
thermal velocity of the
molecule of  $i$ component. The
 condensation coefficient is
included into the effective thermal velocity.

On the base of equations
\begin{equation}\frac{W^{+}_{i}}{W^{-}_{i}} = exp(-\frac{\partial
F(\{\nu_{i}\})}
{\partial \nu_{i}})
\end{equation}
we define inverse flows $W^{-}_{i} $. Then the velocities
can be found as
\begin{equation}
\frac{d \nu_{i}}{dt} = W^{+}_{i} - W^{+}_{i}
\end{equation}

After neglecting $2a v_{l\ i}(\sum_{j} \nu_{j}v_{l\ j})^{-1/3}/3$
in comparison with
 $b_{i}$ for supercritical
embryous in expressions for free energy derivatives we have
\begin{equation} \label{xxx}
\frac{d\nu_{i}}{dt} =
[ \zeta_{i} +1 - \mu_{i}
f_{i}(\mu_{i})]
\frac{(\sum_{j} \nu_{j} v_{l\ j} )^{2/3}}
{\tau_{i}}
\end{equation}

\section{Stationary rate of nucleation}

After instantaneous creation of supersaturation the external influence
interrupts.
 Near-critical region is extracted by the condition
\begin{equation}
\mid F(\{\nu_{i}\}) - F_{c} \mid \leq 1
\end{equation}
where index "$c$" here and below marks the values at saddle point. The time of
relaxation in near-critical region can be above estimated. Let us transmit
by linear transformation to the set  of $\nu_{*\ i}$ which reduces
the square form of the free energy to the the sum of squares and
every $\nu_{*\ i}$ has the same scale as $\nu_{i}$ has. Thanks to the
fact that the set $\{\sum_{j} \nu_{j} v_{l\ j} ,\ \{\mu_{i}\}_{i=1}^{i=n-1}\}$
 diagonalises the square form of free energy it is easy to do.
Then we have
\begin{equation}
t_{rel} = \frac{(diam \nu_{c})^2}{W^{+}_{c}}
\end{equation}

where
\begin{equation}
diam \nu_{c} = [2/(\sum_{j}\frac{\partial^{2}F}
{\partial \nu^{2}_{*\ j}}) )]^{1/2}
\end{equation}
$$
diam \nu_{c}
\sim
perim \nu_{c}
$$
$$
perim \nu_{c}= \sum_{j}[
2/(\frac{\partial^{2}F}{\partial \nu^{2}_{*\ j}})]^{1/2}
$$
\begin{equation}
W_{c} = min[W^{+}_{i\ c}]
\end{equation}
and derivatives are taken at the saddle point  of free energy. One can obtain
this estimate
by sequential considering of diffusion over the first and over the second
variable.
Long tails of the near-critical region
along the diagonals orf rectangular with coordinates
$$ \nu_{*\ i\ c} \pm (2/(\frac{\partial^{2}F}{\partial \nu^{2}_{*\ i}}))^{1/2}
\ \ \ \ \
$$
are omitted. After
relaxation in near critical region the rate of embryos formation is equal
\begin{equation} \label{flow}
J_{s}(\{ \zeta_{a} \}) = Z [
\sum_{j} \Psi_{j} exp(-\Delta_{j} F_{c})
]
\end{equation}
Here
\begin{equation}
\Psi_{j} = \frac{n_{j}}{\sum_{i}n_{i}}
\end{equation}
\begin{equation}
 \Delta_{j} F_{c} = F(\nu_{j\ c}, \nu_{b\ c}) + \ln(n_{j}v_{l\ j})
\end{equation}
 $Z$ is Zeldovitch [3] factor. It is extracted by the fact that all
other terms exept $Z$ appear from equilibrium distribution.
The value of additions to free energy is directly connected
with reconsideration made by Lothe and Pound [6].
It can be corrected in the manner of Reiss.

We assume that in near-critical region there
exists the quasistationary state of embryos.
Then the rate of nucleation is equal to stationary one defined by (\ref{flow}).
 This
situation takes place if the time of characteristic violation of near-critical
state $\Delta t$ strongly exceeds $t_{rel}$.
This fact will be proved on the base of detailed results  for $\Delta t$
 in frames of (\ref{4}).

 For $\nu_{i\ c}$  we
have the system of equations
\begin{equation}
b_{i} = \frac{2 a}{3} (\sum_{j}v_{l\ j}\nu_{j\ c})^{-1/3}v_{l\ i}
\end{equation}
which have unique solution in ideal solution approximation.
When we use  more complicated approximations
 then many nucleation channels can appear.
In every channel the embryo must overcome only one saddle point [21].
So the generalization of the theory is easy.

We can rewrite the
expression for nucleation rate in the folowing ordinary form
\begin{equation}
J_{s}( \{ \zeta_{a} \}) = Z^{+}  exp(- F_{c})
\end{equation}
with renormalized Zeldovitch factor
\begin{equation}
Z^{+} =
Z [\sum_{j} \frac{n_{j}}{\sum_{i}n_{i}}
(\zeta_{j}+1)
\frac{n_{jj}}{v_{l\ j}}]
\end{equation}

This factor is rather smooth function of supersaturations in comparison with
exponential factor in expression for nucleation rate. The sharp dependence of
$  exp(- F_{c})$ on supersaturations can be observed from expressions for
$dF_{c}/d \zeta_{i}$. For
$dF_{c}/d b_{a}$ we have
\begin{eqnarray}
\frac{dF(\nu_{a\ c}, \nu_{b\ c})}{d b_{a}} =
\frac{\partial F(\nu_{a}, \nu_{b})}{\partial b_{a}} \mid_{c} +
\frac{\partial F(\nu_{a}, \nu_{b})}{\partial \nu_{a}} \mid_{c}
\frac{\partial \nu_{a}}{\partial b_{a}}   +
 \sum_{b \neq a}
\nonumber \\
\frac{\partial F(\nu_{a}, \nu_{b})}{\partial \nu_{b}} \mid_{c}
\frac{\partial \nu_{b}}{\partial b_{a}}   + \sum_{b \neq a}
\frac{\partial F(\nu_{a}, \nu_{b})}{\partial b_{b}} \mid_{c}
\frac{\partial b_{b}}{\partial b_{a}}
\end{eqnarray}

The derivative in r.h.s. of the last equation is partial one
because it doesn't depend on the variation of the supersaturation
of the other component.
Meanwhile the $F_{c}$ dependence on $\zeta_{i}$ through
$\nu_{i\ c}$ is taken into account. We can notice that
\begin{equation}
\frac{\partial F(\{\nu_{a}\})}{\partial \nu_{a}} \mid_{c} = 0
\end{equation}
\begin{equation}
\frac{b_{a\ c}}{v_{l\ a}} = \frac{b_{b\ c}}{v_{l\ b}}
\end{equation}
and then
\begin{equation}
\frac{\partial F(\{\nu_{a}\})}{\partial b_{a}} \mid_{c} =
- \frac{(\sum_{b} v_{l\ b} \nu_{b})}
{v_{l\ a}}
\end{equation}
\begin{equation}
\frac{\partial F(\{\nu_{a}\})}{\partial \zeta_{a}} \mid_{c} =
- \frac{\sum_{b} v_{l\ b} \nu_{b}}
{v_{l\ a}(\zeta_{a}+1)}
\end{equation}

In capillarity approximation  inequality (\ref{4}) is valid and
the sharp dependence on supersaturations is obvious.

{}From the last equation it follows that during the period of essential
formation
of the embryous the relative variation of supersaturation is small. Namely,
\begin{equation}
\frac{\Delta_{1} \zeta_{i}}{\zeta_{i}} \leq \Gamma_{i}^{-1}
\end{equation}
where
\begin{equation}
\Gamma_{i} =
\frac{\zeta_{i}}{\zeta_{i}+1} \mid \frac{dF_{c}}{db_{i}}\mid
\end{equation}

\section{Concentration of the solution in droplets}

Now let us define the concentration of droplets at this
period.
In this section we shall write $\zeta_{i}$ instead of $\zeta_{i}+1$.
The dynamic equations for ${\mu_{i}}$ can be written in the
following manner
\begin{equation}
\frac{d\mu_{i}}{dt} = \frac{(\sum_{b} \nu_{b} v_{b})^{2/3}}{\sum_{j}\nu_{j}}
[
\frac{\zeta_{i}-\mu_{i}f_{i}(\mu_{i})}{\tau_{i}} -
\mu_{i}
\sum_{j}
\frac{\zeta_{j}-\mu_{j}f_{j}(\mu_{j})}{\tau_{j}}
]
\end{equation}
As far as we don't express the term of the last equation through $\mu_{i}$,
we work now in the set of variables
$\{\mu_{i}\}_{i=1}^{n-1},(\sum_{b} \nu_{b} v_{b})^{2/3}/\sum_{j}\nu_{j}$ [22].
The stationary solutions are obtained from
\begin{equation}
\frac{\zeta_{i}-\mu_{i}f_{i}(\mu_{i})}{\tau_{i}} =
\mu_{i}
\sum_{j}
\frac{\zeta_{j}-\mu_{j}f_{j}(\mu_{j})}{\tau_{j}}
\end{equation}
The values of $\mu_{i}$ aren't independent but satisfy the obvious restriction
\begin{equation} \label{*}
\sum_{i} \mu_{i} = 1
\end{equation}
So we have to use Lagrange method. Let us measure time $t$ here in units of
$(\sum_{b} \nu_{b} v_{b})^{2/3}/\sum_{j}\nu_{j}$. We obtain potential $U$ from
\begin{equation}
\frac{d\mu_{i}}{dt} = - \frac {\partial U } { \partial \mu_{i}}
\end{equation}
The function $U$ exists due to homogeneous character of restricion (\ref{*}).
The equation of conditional extremum for $U$ will be the following one:
\begin{equation}
\frac{\partial U }{\partial \mu_{i}} - \alpha \frac{\partial}{\partial \mu_{i}}
[
\sum_{j} \mu_{j} - 1
]
= 0
\end{equation}
with arbitrary $\alpha$. This equation leads to
\begin{equation} \label{**}
\alpha +
\frac{\zeta_{i}-\mu_{i}f_{i}(\mu_{i})}{\tau_{i}} =
\mu_{i}
\sum_{j}
\frac{\zeta_{j}-\mu_{j}f_{j}(\mu_{j})}{\tau_{j}}
\end{equation}
The value of $\alpha$ must be chosen according to (\ref{*}). Let us introduce
\begin{equation}
\delta_{i} =
\frac{\zeta_{i}-\mu_{i}f_{i}(\mu_{i})}{\tau_{i}}
\end{equation}
and
\begin{equation} \label{***}
\Delta = \sum_{j} \delta_{j}
\end{equation}
{}From (\ref{**}) we see that
\begin{equation} \label{TTTT}
\frac{\alpha}{\mu_{i}} +
\frac{\zeta_{i}-\mu_{i}f_{i}(\mu_{i})}{\mu_{i} \tau_{i}} = \Delta
\end{equation}
is invariant for every $i$. In the approximation of an ideal
solution  equation(\ref{**})
can be rewritten as
\begin{equation}
\alpha + \delta_{i} = (\zeta_{i} - \delta_{i} \tau_{i}) \Delta
\end{equation}
and gives an expression for $\delta_{i}$ through $\Delta$. Substitution
of this expression into
(\ref{***}) leads to
\begin{equation} \label{x}
\Delta = \sum_{i} \frac{\zeta_{i} \Delta - \alpha}{1+\tau_{i}\Delta}
\end{equation}
Condition (\ref{*}) also must be expressed through $\Delta$:
\begin{equation} \label{y}
\sum_{i}
\frac{\zeta_{i} + \alpha \tau_{i}}{1+\tau_{i}\Delta} = 1
\end{equation}
Equations (\ref{x}) and (\ref{y}) coincide when $\alpha=0$. So
$\alpha=0$ ensures extremum and the condition on $\Delta$ is reduced to
\begin{equation}
1= \sum
\frac{\zeta_{i}}{1+\tau_{i}\Delta}
\end{equation}
Due to $\zeta > 0$ , $\tau_{i} > 0$ the uniqueness of $\Delta$ is obvious.
The last equation is an ordinary algebraic equation of power $n$. As far as in
solution ordinary there are no more than four or five components,
this equation can be
easily solved analytically. This equation can be also solved by
iteration procedure.
For rather intensive process of condensation it is necessary to have
at least one supersaturation many times geater than unity. This value(-es) can
be regarded as the leading parameter in iterational procedure.

{}From the consideration of the limiting case one can
prove that this extremum is stable one. On the base of $\Delta$
concentrations can be reconstucted with the help of (\ref{TTTT}). Only there we
shall use condition $\mu_{i} > 0$.
The relaxation  characteristic time can be found from
\begin{equation}
t_{rel} \approx
[
\frac{(\sum_{b} \nu_{b} v_{b})^{2/3}}{\sum_{j}\nu_{j}}
[
-\frac{1}{\tau_{i}}(f_{i}(\mu_{i}) + \mu_{i} \frac{df_{i}}{d\mu_{i}}) -
 \sum_{j}\frac{\zeta_{j}-\mu_{j}f_{j}}{\tau_{j}} -
\mu_{i} \frac{f_{i}}{\tau_{i}} -
\frac{\mu_{i}^2 }{\tau_{j}} \frac{df_{i}}{d\mu_{i}}
]
]^{-1}
\end{equation}
in approximation of independent $\mu_{i}$.
Here we must put
all values to some characteristic values at the period of droplets formation
and take this expression at the stationary stable value of $\mu_{i}$.
They must be taken from futher considerations or by simple estimates.

Under the conditions of cappilarity
approximation applicability for critical embryo description the following
ineqality
\begin{equation} \label{39}
t_{rel} \ll \Delta_{1} t
\end{equation}

is valid where $\Delta_{1} $
marks the variation of the magnitudes during
the period of droplets essential formation.
After obtaining the value of $\Delta_{1} t$ we can prove
this inequality analytically by the simple substitution.

\section{Evolution equation}

As the result of the previous section we can see that
\begin{equation}
\nu_{a} = \mu_{a} \sum_{b}\nu_{b}
\end{equation}
and
\begin{equation}
\frac{d\nu_{a}}{dt} = \frac{d \nu_{b}}{dt}
\frac{\mu_{a}}{\mu_{b}}
\end{equation}
with constant value of $\mu_{i}$.
At constant supersaturations we can integrate the law of growth
\begin{equation}
\nu_{a} = \nu_{a}(t(\nu_{a})) \equiv
(\zeta_{a}+1 - \mu_{a} f_{a} )^3
(\sum_{j} \frac{v_{l\ j} \mu_{j}}{\mu_{a}})^2
\frac{(t-t(\nu_{a}))^3}{(3\tau_{a})^3}
\end{equation}
where $t(\nu_{a})$ is the moment of droplet formation
and $\mu_{i} $ marks some stable value for arbitrary form of activity
coefficients.
There remains a question about correspondance between the members of the set
$\mu_{c\ i}$ and the members of the set $\mu_{i}$. This question can be solved
on the base of detailed analysis of transition from near-critical region to
supercritical region which forms the matter of separate publication (see [22]).
In the case when $f_{i}=1$ this question doesn't appear.

The values
\begin{equation}
\rho_{a} = \nu_{a}^{1/3}
\end{equation}
grow with constant velocity and the
evolution can be expressed in these values. We have simple connection
\begin{equation}
\rho_{a} =  (\frac{\mu_{a}}{\mu_{b}})^{1/3} \rho_{b}
\end{equation}

Let us construct the condensation equations system. Initial values of
supersaturations are marked by $\Phi_{a}$.
We shall use in this section the lateral "$f$" for distribution functions.
Together with full distribution function $f(\{\rho_{a}\}, t)$ we shall
introduce
\begin{equation}
f_{a}(\rho_{a},  t) = \int_{0}^{\infty} f(\{\rho_{a}\}, t)
\Pi_{b \neq a} d\rho_{b}
\end{equation}
Equations of evolution
can be written for supercritical embryos in the following form
\begin{equation}
\frac{\partial
 f(\{\rho_{a}\},  t) }
{\partial t} =-\sum_{b}
\frac{\partial
 f( \{ \rho_{b} \} , t) }
{\partial \rho_{b}}
\frac{d\rho_{b}}{dt}
\end{equation}
\begin{equation}
\frac{\partial
 f_{a}(\rho_{a},  t) }
{\partial t} =-
\frac{\partial
 f_{a}(\rho_{a},  t) }
{\partial \rho_{a}}
\frac{d\rho_{a}}{dt}
\end{equation}
We introduce variables
\begin{equation}
z_{a} = \nu_{a}(t(\nu_{a})=0)
\end{equation}
So we have
\begin{equation}
z_{b} = \frac{z_{a}\mu_{b}^{1/3}}{\mu_{a}^{1/3}}
\end{equation}
We can write solutions of evolution equations in the following form
\begin{equation}
f_{a}(\rho_{a},t) = f^{+}(x_{a})
\end{equation}
\begin{equation}
f_{b}(\rho_{b},t) =
f_{a}((\frac{\mu_{a}}{\mu_{b}})^{1/3}\rho_{b},t)
(\frac{\mu_{a}}{\mu_{b}})^{1/3}
\end{equation}
where $f^{+}$ is some function of variable
\begin{equation}
x_{a} = z_{a} - \rho_{a}
\end{equation}
We can present $f_{b}$ as some function of variable
\begin{equation}
x_{b} = z_{b} - \rho_{b}
\end{equation}
We shall also introduce
\begin{equation}
\rho = (\sum_{a}\rho_{a}^{3} )^{1/3}
\end{equation}
\begin{equation}
z = (\sum_{a} z_{a}^3 )^{1/3}
\end{equation}
\begin{equation}
x=z-\rho
\end{equation}
All functions of $t$ are functions of $x_{a},x$.
With the help of inequality
\begin{equation} \label{62}
\Delta_{1} z_{a} \gg ( \nu_{a\ c})^{1/3}
\end{equation}
and inequality (\ref{39}) the form of $f^{+}(x)$
 is obtained from boundary conditions for
$f_{a}(\rho_{a},t)$
 which goes to quasi-stationary distribution at small $\rho_{a}$ for all $a$.
This quasi-stationary
distribution corresponds to stationary intensity of droplets
formation $J_{s}$. Then we have
\begin{equation}f^{+}(x_{a}) =
f_{a\ s}(\{ \zeta_{a}(x_{a}) \} )
\end{equation}
where
\begin{equation}
f_{a\ s}(\{ \zeta_{a}(x_{a}) \} )
= \frac{J_{s}(\{ \zeta_{a} \} )}{d\rho_{a}/dt}
\end{equation}

\section{The system of equations of  condensation}

We shall close the condensation equations system
by the balance equations for the substance:
\begin{equation}
\Phi_{i} = \zeta_{i} + g_{i}
\end{equation}
where
\begin{equation}
g_{i} = \frac{1}{n_{ii}}
\int_{0}^{\infty} d\rho_{i} \rho_{i}^3 f_{i}(\rho_{i},t)
\end{equation}
For $g_{i}$ we have the following equations
\begin{equation}
g_{a} = \frac{1}{n_{aa}}
\int_{0}^{z_{a}} dx_{a} (z_{a} - x_{a})^3 f_{a\ s}(\zeta_{a}, \zeta_{b})
\end{equation}
Then from balance equations it follows
\begin{equation} \label{69}
\frac{\Phi_{a} - \zeta_{a}}{n_{bb} \mu_{a}} =
\frac{\Phi_{b} - \zeta_{b}}{n_{aa} \mu_{b}}
\end{equation}
During the period of essential droplets formation where $\zeta_{a}$ is near
$\Phi_{a}$ the following approximation
\begin{equation} \label{68}
f_{s\ a} ( \{ \zeta_{a} \} ) =
f_{s\ a} ( \{ \Phi_{a} \} ) exp(\sum_{i} \Gamma_{i}( \{ \Phi_{a} \} )
\frac{\zeta_{i}-\Phi_{i}}{\Phi_{i}})
\end{equation}
is valid. On the base of (\ref{69}) we come to equation
\begin{equation}
f_{s\ a} ( \{ \zeta_{a} \} )=
f_{s\ a} ( \{ \Phi_{a} \} ) exp(\Gamma_{i}^{+}( \{ \Phi_{a} \} )
\frac{\zeta_{a}-\Phi_{a}}{\Phi_{a}})
\end{equation}
where
\begin{equation}
\Gamma^{+}_{a}(\{\Phi_{i}\}) =
 \sum_{b} \Gamma_{b}(\{\Phi_{i}\})
\frac{\Phi_{a} n_{aa} \mu_{b}}{\mu_{a} \Phi_{b} n_{bb}}
\end{equation}
Now we can obtain from $a$ component balance equation the closed equation
\begin{equation}
\zeta_{a}(z) =
\Phi_{a} -
\frac{f_{s\ a}(\{\Phi_{i}\})}{n_{aa}}
\int_{0}^{z_{a}}
(z_{a}-x_{a})^3
exp(\Gamma^{+}_{a} \frac{\zeta_{a}-\Phi_{a}}{\Phi_{a}}) dx_{a}
\end{equation}
This equation is analogous to the balance equation in one
component theory and can be solved by methods [14].
So we needn't to investigate it here.
All parameters of the period of essential formation of droplets
are obtained now.

{}From
\begin{equation}
\frac{d^2 g_{i}} {d t^2}
\sim
\frac{d^2g_{i}}{dx^2_{a}}
> 0
\end{equation}
and as far as
$$ f_{a}(x) \le f_{*\ a}$$
we have
$$
 (\frac{ 4  n_{aa} \Phi_{a} }
{f_{*\  a} \Gamma^{+}})^{1/4}
\frac{\tau_{a}}{\Phi_{a}+1 }
\le
\Delta t \le (\frac{ 4  n_{aa} \Phi_{a} e}
{f_{*\  a} \Gamma^{+}})^{1/4}
\frac{\tau_{a}}{\Phi_{a}}
$$
This inequality allows us to justify (\ref{39}),
(\ref{62}) and quasistationary state of near-critical
embryos.

The validity of the linearization of the free energy can be proved
taking into account that among supersaturations there are those ones
which are rather great in comparison with unity (it is necessary for
intentive process of nucleation). So in very rought approximation we can
keep only these supersaturations and obtain the expression which resembles
one dimensional case.

\section{Further evolution}

In the description of the further evolution
it isn't necessary to obtain intensity of
droplets formation. So it is sufficient to know supersaturatioins
with small relative error. This fact allows us to use monodispersious
approximation for droplets size spectrum. At the initial period of condensation
when $\zeta_{a} = \Phi_{a}$
the velocities of $\rho_{a}$  growth are independent from their
values. The form of spectrum
doesn't change and the spectrum moves as a whole along $\rho$
axis. Soon it becomes monodispersious.
At this period the
role of spectrum in
the balance of substance is negligible and we can spread the
monodispersious approximation on this period also.
Certainly, the intensity of droplets formation is already known.
We shall consider this period in situation $f_{a}=1$. All
generalizations are rather obvious.

During the time the supersaturations falls. We shall investigate this
evolution.
Let us transform to variables $\mu_{i}$ and $\gamma$
\begin{equation}
\gamma = (\sum_{b}\nu_{b}v_{l\ b})^{1/3}
\end{equation}
On the base of (\ref{xxx}) we see that
\begin{equation} \label{81}
\frac{d\nu_{a}}{dt} =
\frac{\gamma^2}{\tau_{a}}(\zeta_{a}+1-\mu_{a})
\end{equation}
Considering $\gamma$ as parameter we see from (\ref{81})  that during
the evolution the spectrum becomes more and more monodispersious.
For $\gamma$ we have equation
\begin{equation}
\frac{d\gamma}{dt} = \frac{1}{3}  \sum_{a} v_{l\ a}
\frac{\zeta_{a} +1 - \mu_{a}}
{\tau_{a}}
\equiv
\lambda
\end{equation}
which shows that
when spectrum is monodispersious
in $\mu_{i}$ the evolution of $\gamma$ is defined
in stable manner. Hence, the spectrum is monodispersious
in $\mu_{i},\gamma$ (or in $\nu_{i}$).

In monodispersious approximation the condensation
 kinetics equations have the following form
\begin{equation} \label{84}
\frac{d\nu_{a}}{dt} =
\frac{(\sum_{b} v_{l\ b}\nu_{b})^{2/3}}
{\tau_{a}}
(\Phi_{a}+1-\mu_{a}
- N
\frac{\nu_{a}}{n_{aa}})
\end{equation}
where N is the total number of droplets.
 The initial conditions are obvious ones:
\begin{equation}
\nu_{a}(t=0) = 0
\end{equation}
The values of $\nu_{a}$  present the coordinates of monodispersious peak.

At first let us consider the case $\zeta_{a} \gg 1$ for all $\zeta_{a}$. Then
\begin{equation} \label{87}
\frac{d\nu_{i}}{dt} =
\frac{\gamma^2}{\tau_{i}}(\Phi_{i} - N \frac{\nu_{i}}{n_{ii}})
\end{equation}

System (\ref{87}) doesn't posess analytical solution.
 The next simplification
is to consider $\gamma$ as known function of time.
The values of $\nu_{a}$ in the expression for $\gamma$
are
considered to be formed at ideal supersaturation. Then
$$\gamma = \lambda(\{\Phi_{a}\})t$$
and (\ref{87}) can be integrated
\begin{equation}
\Phi_{i} - N\frac{\nu_{i}}{n_{ii} } =
\Phi_{i}
exp(-N\frac{\lambda^{2}t^3}{3 n_{ii} \tau_{i}})
\end{equation}

The errors can be estimated  by comparison with precise solution
in one dimensional case.
So we have the maximum error in
$\rho$ to be equal 0.047. In one dimension
case this approximation is also important because it allows
to inverse $t$ dependence on $\rho$.

When supersaturations are small in comparison with their initial values we can
construct asymptotes with the help of
methods given in [16].

The deviation of $v$ from $1$ (not the deviation of $n_{aa}/n_{bb}$ from $1$)
leads to exhausting of one component.  We can see that for $\zeta_{a},\zeta_{b}
\gg 1$:
\begin{equation}
\frac{dg_{a}}{dt} \approx \frac{n_{bb} \tau_{b} \zeta_{a} }
{ n_{aa} \tau_{a} \zeta_{b}}\frac{dg_{b}}{dt}
\approx
\frac{\zeta_{a}}{\zeta_{b}} \frac{dg_{b}}{dt}
\end{equation}
The last approximative equality
is valid when thermal velocities (i.e. masses of
molecules) and partial volumes have one and the
same power for different components.
So the consumption of different
components occurs practically
simultaneously and there is no further essential formation of droplets.

\section{References}
\par
1.Volmer, M., Z.Phys.Chem. vol.25, p.555 (1929)
\par
2.Becker, R. and Doering, W., Ann.Phys. vol. 24, p.749 (1935)
\par
3.Zeldovitch,J.B., J.Exp.Theor.Phys. (USSR) vol.24, p.749 (1942)
\par
4.Frenkel J., Kinetic Theory of liquids, Oxford University Press, New York,1977
\par
5.Kramers, H., Physica vol.7, N 4, p.284 (1940)
\par
6.Lothe J. and Pound, G.M., J.Chem.Phys. vol.36, 2082 (1962)
\par
7.Reiss, H., in A.C.Zettlemeyer (ed.), Nucleation Phenomena, Elvier New York,
1977
\par
8.Reiss, H., Katz, J.L., and Cohen, E.R., J.Chem.Phys. vol.48, p.5553 (1968)
\par
9.Feder, J., Russel, K.C., Lothe, J., amd Pound, G.M., Adv. in Physics
vol.15, p.111 (1966)
\par
10.Hung, C.H., Katz, J.L., and Krasnopoler M.J., J.Chem.Phys. vol.90, p.1856
(1989)
\par
11.Reiss, H., Tabazadeh, A., and Talbot, J., J.Chem.Phys. vol.92, p.1266
(1990)
\par
12.Wakeshima, H., J.Chem.Phys. vol.22, p.1614 (1954)
\par
13.Raiser, Yu.P., Journ.Exper.Teor.Phys. (USSR) vol.37, p.1741 (1959)
\par
14.Kuni F.M., Grinin, A.P., and Kabanov, A.S., Colloid.J.
(USSR) vol.46, p.440 (1984)
\par
15.Grinin, A.P. and  Kurasov, V.B., Vestnik Leningrad Univers. (USSR), physics
chemistry, Issue 2, p.100 (1984)
\par
16.Kuni, F.M., The kinetics of the condensation under the dynamical conditions,
Kiev,1984, 65p. Preprint of the Institute for Theoretical Physics IPT-84-178E
\par
17.Kuni, F.M., Grinin, A.P., and Kurasov, V.B., in Mechanics of inhomoheneous
liquids (USSR), Novosibirsk, 1985, p.86
\par
18.Grinin, A.P., Kuni, F.M., and Kurasov, V.B., Colloid.J. (USSR)
vol.52, p.430; p.437; p.444 (1990)
\par
19.Reiss, H., J.Chem.Phys. vol.18, p.840 (1950)
\par
20.Stauffer D., Aerosol Sci. vol.7, p.319 (1976)
\par
21.Kuni, F.M., Kurasov, V.B., Djikaev, Yu. Sh., and Melikhov A.A.,
Thermodynamics and kinetics of binary nucleation, Preprint Freie Universitat
of Berlin, FUB-TKM 12-90, 52p.
\par
22.Kurasov, V.B., The kinetics of the binary condensation in the conditions
of instantaneous creation of supersaturation, VINITI (USSR), N 4440-B90,
from 27.06.1990, 48 p.
\par
23.Kurasov, V.B. and Kurasov P.B. The kinetic description of binary nucleation,
Preprint Manne-Siegbahn Institute for Atomic Physics of Stockholm University,
MSI 93-3, 90p.
\par
24.Kurasov, V.B., Kinetics of the binary condensation in the dynamical
conditions, VINITI (USSR), N 4439-B90, from 26.06.19990, 45p.
\par
25.Lifshitz, I.M. and Slyozov, V.V., J.Exp.Theor.Phys.(USSR) vol.35, p.479
(1958)
\par
26.Lifshitz, I.M. and Slyozov, V.V., J.Phys.Chem.Solids vol.19, p.35
(1961)

\end{document}